# Economics for the Global Economic Order: The Tragedy of Epic Fail Equilibria

Shiro Armstrong and Danny Quah[1]

October 2023

**Abstract**

This paper casts within a unified economic framework some key challenges for the global economic order: de-globalization; the rising impracticability of global cooperation; and the increasingly confrontational nature of Great Power competition.  In these, economics has been weaponised in the service of national interest.  This need be no bad thing.  History provides examples where greater openness and freer trade emerge from nations seeking only to advance their own self-interests.  But the cases described in the paper provide mixed signals.  We find that some developments do draw on a growing zero-sum perception to economic and political engagement.  That zero-sum explanation alone, however, is crucially inadequate.  Self-serving nations, even when believing the world zero-sum, have under certain circumstances produced outcomes that have benefited all.  In other circumstances, perfectly-predicted losses have instead resulted on all sides.  Such lose-lose outcomes—epic fail equilibria—generalize the Prisoner's Dilemma game and are strictly worse than zero-sum.  In our analysis, Third Nations—those not frontline in Great Power rivalry—can serve an essential role in averting epic fail outcomes.  The policy implication is that Third Nations need to provide platforms that will gently and unobtrusively nudge Great Powers away from epic-fail equilibria and towards inadvertent cooperation.

[1] The authors are at, respectively, Crawford School of Public Policy, The Australian National University; and Lee Kuan Yew School of Public Policy, National University of Singapore.  They are grateful for helpful discussions with Selina Ho, Ashutosh Thakur, Sam Hardwick and participants at the Pacific Trade and Development Forum conference in Seoul, Sep 2023.

# 1. Introduction

It is a given that world order, the complex web of relations between nations, is in the midst of disruption. Globalisation—the economic dimension to that order—might not be disappearing but it is certainly changing.  Cooperation to address global challenges is visibly becoming scarcer, even as such challenges become ever more pressing.  And Great Power competition, which once drove nations to make themselves more productive, has taken the approach of seeking to contain rivals rather than self-improvement.

This paper provides an economics-based account of the forces at work in these disruptions. This Introduction summarises key ideas.  The remainder of the paper provides historical and institutional detail.

Global disruptions can arise from exogenous disturbances or they can emerge endogenously from within the system.  This paper focuses on the latter, i.e., on endogenous disruptions.  In analysing these, a reasonable first step might be to conjecture that the world has become zero-sum, that every nation now fears how it loses what others gain. Hence, nations are behaving towards each other in a way that is more confrontational and less constructive. However, the reasoning we develop and the evidence we provide suggest this cannot be the main explanation.

In many current endogenous disruptions, all parties lose.  Examples include variants of geopolitical rivalry that might lead to military confrontation.  In a zero-sum game, by contrast, no matter how much is lost elsewhere in the system, someone always wins.  But the disruptions that matter in the world today are manifestly not zero-sum.  They are worse.

Psychologically, perceptions might indeed be increasingly zero-sum.  But positive-sum improvement to world order is everywhere obvious and feasible for the simple reason that global crises are not in short supply.  The world continues to see urgent need for global public goods, in contrast to individual observers' psychological perception narrowing otherwise to zero-sum.  Some parts of current geopolitical narrative show clear understanding of this, from which Pareto-improving steps ought to emerge through clear-headed international leadership.  Reality is not zero-sum, and if out-of-kilter perceptions were the only barrier, a set of strategies can certainly be devised to advance everyone's well-being.  Yet in many cases, these positive strategies have not eventuated.

This paper suggests instead that rather than being zero-sum, more and more current international situations are converging towards what we call **epic fail** outcomes, i.e., equilibria where everyone ends up worse off. These emerge even when every nation seeks only to advance its own self-interest: their actions end up undermining their own well-being.[2] It is, of course, right and appropriate that nations seek to improve their well-being. What is tragic, however, is when such behaviour leads to a result strictly worse than zero-sum, and everyone loses.

This paper documents the emergence of epic fail outcomes in the global economic order. Again, we suggest that it is this, rather than zero-sum, that now characterizes the direction that Great Powers are taking the international system.

The payoff to our analysis is its simple and direct policy implication. Third Nations—those not in the frontline of Great Power rivalry—will need to exercise agency to shift the international system away from epic fail equilibria. In the process Great Powers have to be nudged gently and unobtrusively. Third Nations should aim to do exactly that. Given the global realities we describe below, explicit agreed-upon collaboration is likely too ambitious a goal. However, inadvertent cooperation might be achievable: This idea is simply that of not letting the perfect be the enemy of the good. Third Nations can create an environment where countries continue to act in their own interest but avoid epic-fail outcomes and instead achieve positive-sum Pareto improvements. Convening a Bandung 2.0 would be, in our view, a reasonable first step to bring about such inadvertent cooperation. Working with a broader conception of security such as comprehensive security, that elevates positive-sum aspects of collective security, can also provide a path forward.

The rest of this paper is organized as follows. Section 2 explicitly relates to world order our ideas on international contestation. Economics most visibly appears in world order from the many international economic institutions—the international financial architecture, the world trading system, foreign investment agreements—that constitute important elements in the international landscape. But our analysis seeks to go further in that it frames the evolution of global order itself in terms of economic dynamics. Section 3 explains why a zero-sum perspective alone is inadequate. It shows how today's Great Power competition in Section 2 leads to outcomes worse than zero-sum.

Section 4 traces a historical arc in the international system, with specific focus on the US: parts of the international system were initially resolutely un-global; then the entire system underwent a half-century of US-led globalisation, and most recently reverted to historical norm.  Section 5 refocuses on the nexus between considerations of economic performance and of national security, and brings out the role of Third Nations.  Section 6 concludes.

---

[2] This is of course what happens in a Prisoners Dilemma situation, but epic fail outcomes are generalizations of Prisoner's Dilemma. For epic fails to emerge, all that is needed is that a Pareto-inferior outcome be an equilibrium and an alternative Pareto-superior outcome be feasible but not equilibrium. For example, over-fishing a common resource or under-investment in the presence of positive externalities are epic fails but not Prisoner's Dilemma. We will return to this below. Quah (2023) presents a fuller discussion distinguishing epic fail outcomes from zero-sum and Prisoner's Dilemma situations.

## 2.  Bringing economics into world order

Economics enters our analysis in two distinct ways. First, international economic institutions are a key ingredient of world order. The health of those institutions co-evolves with the temperature of world order.  Second, world order itself evolves through economic forces.  In our analysis, those forces are identified as perturbations in supply and demand for world order.

Of all international connections, none is more consequential than economic relations, i.e., the patterns of and attitudes towards trade, investment, and other flows of factors of production and consumption across nations. The centrality of economic relations makes international economics both support structure for geopolitical stability in good times and convenient target for cross-national disagreement in bad.  Global economic order is thus bi-directionally intertwined with national security and other dimensions of world order. This deep coupling between geopolitics and international economics can make for either a virtuous or vicious cycle.

Favorable geopolitical conditions induce positive economic relations, and upbeat economic performance in turn reinforces propitious geopolitics.  Good times, therefore, strengthen through co-movement in security and economic relations, manifesting in a virtuous cycle. Conversely, however, when geopolitical conditions turn unfavorable, economic connections come under attack from narrowing nationalist sentiment and, consequently, worsen already adverse international relations. This time, it is a vicious cycle. Geopolitical rivalry emerges; economics is weaponized; national security concerns mount and geopolitical rivalry sharpens yet further.  When vicious-cycle effects take over, economic prosperity can too easily be under-emphasised in policymaking, not least when national security concerns heighten.

Global economic relations thus co-evolve closely with geopolitics and concerns on national security, even if the nexus between them varies over time. No subset of individual variables is causal for the others; all are jointly determined.

Good times, or virtuous-cycle episodes, allow nations to enjoy positive-sum game payoffs, where economic win-win outcomes are the norm. On the other hand, bad times, or vicious-cycle episodes, drive nations towards viewing relations with others as a zero-sum game, or worse, towards epic fail type behaviour.

World order, therefore, follows an endogenous dynamically evolving trajectory. What determines that trajectory? We hypothesise that the endogenous arc of world order is jointly determined by actors on both a supply side and a demand side for world order.[3]

In our reasoning, certain services are performed efficiently by big-nation Great Powers: those services include establishing and enforcing international rules of the game; resolving cross-country disputes where diplomacy has failed; acting as international consumer and lender of last resort; stabilizing the international financial system through, among other things, providing a trusted financial architecture and a world reserve currency; and so on. All these activities are characterized by dramatic increasing returns to scale and are thus prohibitively costly for small states. On the other hand, Great Powers find it in their interest to provide some quantity of global public goods but choose to do so only to a level optimal for themselves but strictly less than globally optimal. Still, it is Great Power actions that shift the needle on such provisions.

In our formulation, those nations able to undertake such economies of scale activities constitute the supply side of world order. All other nations form the demand side who consume these services and somehow, sufficiently or insufficiently, compensate the providers of those services. In the current geopolitical setting, the Great Powers are typically considered to be two, China and the United States. So it is these two, engaged in geopolitical rivalry, that others look to as potential providers of different features of world order. All other nations are Third Nations, constituting the demand side of world order.

In much popular and academic discussion on geopolitics, because what we call the demand side comprises smaller states, Third Nations are viewed as at best price-takers and, by implication, cannot affect outcomes. We disagree with this. In economics price-taking behaviour is completely consistent with equilibrium fluctuating due to shifts in the demand schedule. Even price-taking Third Nations, therefore, have agency. What emerges in equilibrium (Figure 1) depends on both demand-side agency and supply-side elasticity.[4]

---

[3] Quah (2019) previously used this description for his preferred new model of power relations for Southeast Asia and the Great Powers.
[4] This is the approach taken in Quah (2023), extending to an equilibrium setting and formalizing ideas in Acharya (2016), Heng and Aljunied (2015), Ingebrietsen et al (2006), Keohane (1982), and Long (2022).

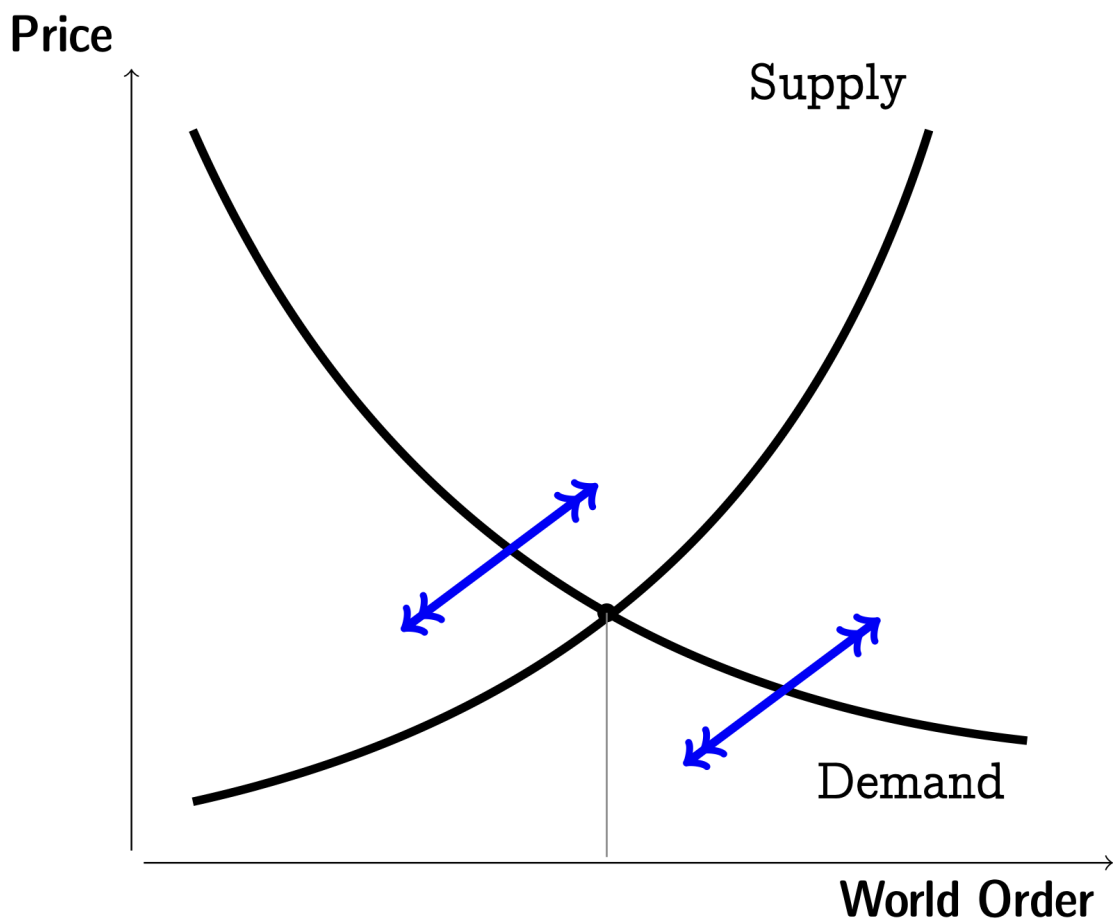

Figure 1. Demand and Supply in World Order. Perturbations in demand shift world order even when supply remains invariant.

---

It has long been implicit in conventional analysis that Great Power rivals offer alternative supply-side settings for world order. Typically not made explicit in such discussions, however, is the demand side. Take, for instance, Cold War rivalry. This was presented to the world by John F. Kennedy as "the long twilight struggle" over "the most ancient cause of all". Kennedy's narrative arrayed democracy and freedom on one side, and totalitarianism and tyranny on the other. While such a proposition might have made sense to those on the supply-side, it was meaningless to everyone on the demand-side as it offered them no agency: All agreed already on wanting freedom; no one wished for tyranny. In this formulation of Cold War rivalry, the demand side is not given freedom of choice.

A second leading example is how in current US-China rivalry, a question often posed to Third Nations is, "Which side will you choose?" (e.g., Lee 2020). While this takes a step beyond Kennedy's original framing and is suggestive of Third Nation agency, it doesn't get very far.

By posing the question to Third Nations as simply one of picking sides—this Great Power or that—such casting subverts the legitimate interests of Third Nations to a narrow issue of simple alignment.[5]

(Our statement is intended to be stark, as obviously both China and the United States have pressured Third Nations to pick sides, sometimes explicitly with carrots and sticks in the form of economic sanctions. Thus, the binary choice characterisation is an over-statement but it points ahead to how the choice set for Third Nations is narrowing.)

In the remaining sections this paper uses this economic lens of supply and demand to discuss the challenges of the global economic order. We will document how Great Powers set the supply curve and thus powerfully help determine the tone and temperature of geopolitics. But supply is only part of the story; demand matters too. So when that supply side is seized by perceptions that drive a vicious cycle dynamic towards epic fail, it is incumbent upon the demand side, Third Nations, to help nudge equilibrium towards better outcomes. The historical record that we review in the sections that follow provides examples where equilibrium is, indeed, influenced by such demand-side actions.

This theoretical reasoning also informs our conclusions: Vicious-cycle effects mean that without appropriate counter action, an insecure and stagnant global economy will emerge as a new normal. This is an epic fail as Pareto-improving cooperative actions are available that can advance the interests of all nations simultaneously.

## 3. Great Power competition leading to an Epic Fail

Competition in general and Great Power competition in particular need be no bad thing. In general, competition drives improvement: Sure, not all rivals on the supply side succeed but everyone on the demand side gains when more is supplied at lower price.

---

[5] It is here where our paper diverges most from, say, Tucker (2022) who also seeks to place Great Power rivalry in an explicit economic and mechanism-design setting. Our paper seeks to present a balanced perspective on Third Nation agency in equilibrium analysis, while Tucker (2022) continues to emphasise the more conventional narrative of power and top-table decision-making. As Quah (2019) notes, that conventional narrative follows the Thucydides dictum that "Great Powers do what they will, the rest of us suffer what we must", whereas Figure 1 is precisely a rejection of that position. Tucker (2002) emphasises values and civilizational contests, and envisions international cooperation among liberal democracies as a way to deal with China and other illiberal states. By contrast, our view is that analysis and history both suggest limits to explicit collaboration so we seek only inadvertent cooperation among Great Powers and Third Nations, symmetrically, through aligning economically-meaningful incentives. Analysis that aims to insert Third Nation agency into a picture of Great Power rivalry cannot, in our reckoning, prioritise civilizational and political values over economic cost and benefit.

How then is today's Great Power competition damaging the international order? We describe some key markers, and show how they relate to our previous discussion on the nexus between economic relations and geopolitics. In short, the competition that the Great Powers are engaged in today is one that is reducing the supply of global order.

China and the US today engage in competition to determine the future contours of the international political and economic system. There is contestation, notably, over Taiwan's political system; the rules of international trade; and leadership in technology (in particular in quantum information science, AI, biotechnology, next-generation communications, semiconductors, green energy; Allison et al, 2021).

Outside these domains of competition, other aspects surface but are not, in our view, central. Containment of China's economic growth is not logically an end goal, but instead an intermediate, enabling objective for the US. This is because a larger, more powerful China will make for a more daunting opponent.

On the other end, China certainly prioritises a successful economy. There is nothing objectionable in that. Ideological control of China's industrial production is not, in our view, an ultimate objective of the Chinese state, but matters to the extent that control is understood to help strengthen the economy as needed at critical moments. The desire for control on the part of the Chinese state is more about security, power and political stability, and economic activity is sacrificed, from time to time, but the economic imperative has remained critical. In our interpretation, the ideas surrounding John F. Kennedy's "long twilight struggle" gain only minimal traction in today's China-US rivalry. US-China competition is not a repeat of the Cold War, a contest over ideology and way of life. It is instead a matter of political and economic dominance.

Competition for global primacy is perhaps more evident from the actions of the United States. But China's rise is coupled with its desire for geopolitical space and international respect, and that increases pressure on its neighbours and the existing system.

Thus, consequential is the set of signals that Third Nations are receiving and to which some are already responding. As previously described, commonly heard in many discussions is how Third Nations are asked to choose sides between the US or China. Many refuse to declare, but their actions reveal their positions. Some larger Third Nations have already openly taken up the US's call for containing China, and have implemented policies accordingly.

Many Third Nations find appealing both China's large, growing market and China's operating principles of national sovereignty and territorial integrity. While those principles conflict with Russia's actions in its invasion of Ukraine, China cannot publicly condemn Russia, something China might have done if it weren't itself a target of containment.

Our conclusion is that in today's Great Power competition, there is no longer a simple Kennedy-type partitioning of sides. Instead, the division is far more complex and multi-dimensional. Complexity, however, complicates the narrative for unifying domestic constituencies. Whether political leadership makes it so, or citizens simplify it thus, the outcome is that today's geopolitical competition comes down to asking which Great Power will be number 1. Ranking is just an alias for all the myriad ways in which nations view gains and losses in relative terms rather than absolute magnitudes (Powell 1991). Competition over ranking makes geopolitical rivalry a zero-sum game. One nation can ascend to number 1 only if the incumbent loses that position. In what we identified earlier as the core issues in US-China rivalry, there are some zero-sum features but there are also positive-sum ones, where both sides can gain. Trade and technological collaboration are examples of the latter.[6]

In contrast, zero-sum competition provides no positive-sum space. The defining feature of zero-sum thinking is that one side wins only when the other loses. In such a situation there is no room for compromise and no room for win-win solutions. Unfortunately, it is a zero-sum situation when nations begin to analyze geopolitics in terms of pure rankings: Who is number 1, who is number 2? Casting the competition in terms of rankings, one competitor rises only when its rival falls. Zero-sum thinking gets baked into the narrative and removes any room for compromise.

A shift in political calculation from looking at absolute advantage to paying attention to relative gains and losses has the same effect as competition over rankings, in that prioritising the relative too locks in zero-sum thinking. Absolute advantage analysis allows win-win outcomes. In contrast, relative gains calculation always produces an equal loss for every gain. A relative-gains perspective makes the world, again, zero-sum.

In our analysis, therefore, an unfortunate zero-sum perspective has set in from the reduction of the complex set of issues in US-China rivalry to the question of which nation will emerge as the lead nation. Moreover, thinking about competition in terms of relative gains and losses has strengthened that tendency towards a zero-sum mindset. Unless this narrative can be overturned, it will progressively edge out possible areas for compromise that could, instead, lead to a healthier, more stable global economy for all.

Thus, it is natural that a first analysis should have led to the hypothesis that the world has now become zero-sum.

---

[6] Posen (2023) discusses the zero-sum beliefs implicit in recent American industrial policy. The roots and arc of zero-sum thinking specific to US residents are explored in Chinoy et al (2023).

However, more deeply, our analysis suggests that that engagement has moved to something even worse than zero-sum. Zero-sum competition captures the sense that compromise is no longer acceptable. However, from the perspective of the global economy, at least in zero-sum competition whatever one side loses, someone else gains. The tragedy is when nations undertake actions that produce losses for everyone in the system. This is then not a zero-sum game. Instead, the leading example of such lose-lose outcomes is the Prisoner's Dilemma game, where win-win cooperation is possible but is simply not chosen in equilibrium. This is what we mean by an epic-fail situation, where everyone can in principle be made better off, but if each nation seeks to advance only their self-interest, in equilibrium everyone ends up worse off instead. We think geopolitical competition has acquired features of zero-sum thinking in some dimensions and epic-fail competition in yet others.

The ideas that underpin this dynamic are usefully made explicit. Economic competition is when a nation attempts to get ahead by increasing its productivity: it invests in schools, encourages research and development, improves public infrastructure, and raises human capital by developing skills and training the nation's workforce. Such actions elevate the nation's economic performance at the same time they improve the situation for the whole world. Security competition, in contrast, is when one nation attempts to get ahead by keeping others down, encircling and containing its rivals, and keeping from them frontier technologies. In security competition, whether the nation ends up absolutely stronger or weaker is secondary: What matters is how the nation does relative to its opponents.

It is tragic when policy thinking ignores the historically-confirmed positive-sum, virtuous circle between security and economics, and instead assumes zero-sum competition. But in a zero-sum contest, at least someone wins. The epic fail tragedy is even worse than a zero-sum outcome. In our current context it is when even with win-win outcomes available, inappropriate geopolitical narratives end up producing equilibria where everyone loses.

Systematizing and extending the analysis, for an epic fail description to apply, two ingredients must be present. First, there must be available outcomes where everyone wins. In the international system, Great Power cooperation would get us to a win-win outcome on mitigating the global climate crisis, driving a clean energy transition, stabilizing the international financial system, efficiently sharing data in the digital economy, and making global supply chains more secure, resilient, and efficient. When win-win is possible, anything less is inefficient and sub-optimal.

The second ingredient for an epic fail is that action based on individual self-advancement, given all others similarly optimize their self-interests, produces an equilibrium where everyone loses. A third ingredient may be a force, like zero-sum psychology, that is preventing cooperation.

It is these ingredients that in game theory drive the Prisoner's Dilemma game. But the current geopolitical situation is worse than a Prisoner's Dilemma outcome. Zero-sum security or strategic competition is suboptimal in security terms: it avoids collective security and makes the world more dangerous.  The reality is that viewing economic engagement as a zero-sum game, weaponizing trade as a result, ends up self-harming, so that the situation transmogrifies into an epic-fail. The game is now no longer merely zero-sum. The epic-fail outcome results in a vicious cycle of economic self-harm that unravels political cooperation and the Capitalist Peace.

The world today is characterized by all these features we have described. The US justifies its use of industrial policy based on the belief that market failure is rife, and that growth alone is not sufficient as a goal for social advancement (Sullivan 2023): negative externalities destroy the environment and improvement across the entire income distribution is important, not just increases in the average. Hence, the US feels it can advance its national security interests and the well-being of its people through industrial policy and controlled international economic frameworks instead of free trade.  China looks to reassert state control over markets and advances a narrative of Common Prosperity whose goal is to rein in excessive income inequality.

To those who live in neither the US nor China, the economic views of the Great Powers sound remarkably aligned.

That these positions inject a note of skepticism on untrammeled free trade is no obstacle to international economic exchange.  In finance, stochastic uncertainty is no obstacle to markets' well-functioning: the relevant returns simply become risk-adjusted returns, rather than pure expectations.  So too international trade can easily factor in adjustments for concerns over resilience, externalities, and income distribution.  The need for assurance on supply should lead to ever greater international trade, with as many other nations as possible, not to closing down to working with just a select number of "friends and like-minded partners".

If the US feels that China has not lived up to an implicit bargain of becoming liberal politically with increasing economic prosperity, China correspondingly feels its geopolitical rival has not maintained the rules of international economic competition.

These considerations suggest to us that the circumstances surrounding current US-China rivalry have now surfaced an epic-fail, not just zero-sum.  In this section, therefore, we have connected our theory of zero-sum competition and epic-fail outcomes to features of geopolitical reality.  We have shown how zero-sum analysis alone is inadequate, and that geopolitics has now pushed into epic-fail territory.

To further support our theory, we now provide an account on how geopolitical competition currently shows features of zero-sum and epic-fail thinking.  What follows is obviously not an exhaustive account of developments over the last century that bear on international economic relations.  It is, instead, an idiosyncratic narrative that brings out our framing of episodic shifts between zero-sum perception, win-win outcomes, and the epic-fail game in the global economy.

## 4. Change and uncertainty in the global economic order

Today, Great Power competition threatens an already weakened international order. But is a continuing downward spiral inevitable? This section argues the answer is No, based on historical evidence viewed through the lens of our model described in Sections 2 and 3.

A century ago the 1929 Wall Street stock market crash precipitated the Great Depression, eventually manifesting as a 15% fall in world GDP. Over the following decade unemployment in the US rose as high as 23%, and in other parts of the world up to 33%. Consumer confidence fell sharply: Despite increased business and government spending in early 1930, the fall in consumer expenditure and exports ended up lowering aggregate demand worldwide.

In every nation the worsening economy encouraged protectionism in the form of "beggar thy neighbor" policies. The reasoning was that if the margin of reward could be shifted away from foreign towards home production, that would help lift the domestic economy and benefit domestic workers and industry.

This thinking reversed the century-long movement towards global free trade in the 1846 repeal of the Corn Laws in Britain, the first globalization from 1870 through the beginning of the First World War, and the League of Nations' 1927 World Economic Conference that had sought to reduce tariffs worldwide.

Following the Wall Street crash, the US government put in place the Smoot-Hawley Tariff Act of 1930, increasing tariffs on tens of thousands of imported goods. It remains debated among economists how much the Act affected US business activity—for reasons of both timing and magnitude. But there is little question that Smoot-Hawley sparked a raft of protectionist measures around the world. Because other nations had trading sectors proportionally larger, the impact on them was worse: this rise in protectionism exacerbated the decline in world GDP[7].

The exigent circumstances of the Great Depression, therefore, led to a U-turn in progress on trade openness that, around the world, many nations up through 1929 had been steadily driving. For the US, however, the Smoot-Hawley Act simply continued long-standing national protectionism. Alexander Hamilton, the first Secretary of the Treasury (1789-1795), considered open trade impracticable, and was a strong proponent of import tariffs and outright import bans. Abraham Lincoln in 1844 said, "Give us a protective tariff and we will have the greatest nation on earth". In 1896 the Republican Party in the US had its party platform describe protectionism as "the bulwark of American industrial independence, and the foundation of development and prosperity." The reasoning, given in the text of the party platform, was that: "This true American policy taxes foreign products and encourages home industry. It puts the burden of revenue on foreign goods; it secures the American market for the American producer. It upholds the American standard of wages for the American workingman; it puts the factory by the side of the farm and makes the American farmer less dependent on foreign demand and price."

---
[7] Irwin (2011) and Mitchener et al. (2022) estimate that the large welfare losses for the United States was from the retaliation, not the Act itself.

Over all this time, from 1846 through the mid 20th century, the idea was constant and prominent that advancing one's self-interest has strong connection with international economic competition. National advancement and economic competition were always intertwined, even if the algebraic sign on the relationship might change over time and across geographies. No nation was indifferent, although over time they might change their minds or disagree with one another. Pre-1929, in Britain, outside of its imperial preferences, and the world as represented in the League of Nations, the push for open trade came hand in hand with the idea that international exchange made one's nation stronger and more prosperous. With the Great Depression, however, the emergence of beggar-thy-neighbour policies made many nations switch tack: the belief grew that a nation would be better positioned if it disengaged from free and open trade. Protectionism became acceptable as the way to advance national interests.

The US maintained throughout at least an invariant view: right from its very founding and well into the Great Depression, the US consistently believed it would be a successful nation by tilting the playing field and not engaging in free trade and open economic competition with others.

This global convergence on protectionism and increasingly fractured relations across countries, is conjectured to be one of the critical drivers for the eventual onset of World War 2 (Hirschman, 1945, Armstrong et al., 2022). However, the Cold War that emerged afterwards turned out to be formative for the world changing again its approach to the relation between national interests and economic openness.

In the 1950s, the US began a profound break from the mercantilism that had characterised its centuries-long stance on international economic relations. Elected to the Presidency in 1952, Eisenhower took greater leadership on US foreign policy than his predecessors, and in 1955 brought the nation into a new international agency, the Organization for Trade

Cooperation, part of the General Agreement on Tariffs and Trade (GATT) that was actively working to lower obstacles to global trade.  The statement released upon Eisenhower's action stated that "failure on the part of the US, the world's greatest trading nation, to join in setting up this organization (OTC) would cause great dismay and disappointment throughout the free world at a time when the Soviet Union is stepping up its foreign economic efforts" (Egan, 1955).

Put differently, the 1950s saw the US turn away from protectionism because it saw international trade as a way to counter the Soviet security threat.  This marked a sea change in the US's approach to international economic relations.  The globalization of the late 20th century and  US support of it actually began with the view that security and economic competition were complementary.

It is helpful to pause here to appreciate the consequentiality of this adjustment in US perspective.  Some insight on this derives from comparing the Eisenhower pivot with another development just prior to it, the so-called San Francisco System (Calder, 2004).

The 1951 US-Japan Peace Treaty provided for an integrated system of political and economic relations in the Pacific region.  Because of where the treaty was signed, the collection of relations set up also became known as the San Francisco System.  One of the driving ideas of the San Francisco System was to provide Japan economic space to become prosperous and thereby a growth engine in the Pacific region.  This put into practice an important lesson from the treatment of Germany after World War 1 (Keynes, 1919).  The US saw this as getting ahead of what John Foster Dulles identified as a tendency in Japan to revert to violent nationalism amid fraught domestic political instability.  Security and economics, therefore, were intended to work together.

However, the differences were profound between the Eisenhower pivot and the San Francisco System.  The Eisenhower pivot concerned a push for the US to join the Organization for Trade Cooperation, a body within GATT.  The move was explicitly multilateral, and foreshadowed the globalization that would emerge over the next half-century.  The Eisenhower pivot attracted widespread attention and approval from GATT and like-minded nations.  It raised US soft power and elevated America in the eyes of others.

In contrast the San Francisco System focused US attention on only selected parts of East Asia and on Japan primarily.  The system presented a package of bilateral arrangements, with multilateralism notably absent (Calder 2004, p. 140).  The San Francisco System provided special dispensation to Japan and presented the view that that nation, with its fragile politics, should not fall back into violent nationalist fascism.  But notwithstanding even that, all major Asian nations, apart from Japan, objected to the San Francisco Peace Treaty (Calder 2004, p. 146).  Under the San Francisco System, (Calder 2004, p. 144) "America accorded

Japan unusually favorable (and highly asymmetrical) trading and investment arrangements, many of them informal, that Europe did not enjoy. These included informal American tolerance of Japanese trade protectionism, as well as of often substantial discrimination against foreign investment. The US likewise insulated Japan from the reparatory demands of its neighbors, in ways that it did not favor the erstwhile European Axis partners, Germany and Italy."

The multilateralism and openness associated with the Eisenhower pivot were therefore notably missing from the San Francisco System.   Once the US turned with Eisenhower's decision, however, the significance of and space for positive economic engagement rose dramatically around the world.

The bilateral hub and spoke security arrangements of the San Francisco System and multilateralism were complements while the United States had primacy and would supply the global public goods necessary for world order.

The Bretton Woods institutions helped to curtail protectionism and separated trade policy considerations from national security and geopolitics (Cooper, 1972a; Cooper, 1972b). Central was GATT, providing the organizational structure for the rules-based international trade order.  GATT grew into the WTO.  These developments were largely underwritten by the US over the past 75 years, following Eisenhower's landmark policy, and critically helped manage the risks from economic engagement. As testimony to GATT-WTO success, outside the Cold War, geopolitics and economics largely de-linked for close to a century.

Under the US-led multilateral economic order, for those countries that had committed to the system even during the Cold War, the conduct of trade policy could to some extent be pursued separately from national security considerations. Participants surrendered the right to use trade levers to exercise political coercion, except in special and unusual circumstances. The GATT allowed trade between countries under agreed multilateral trade rules that largely quarantined them from geopolitics. There were, of course, economic and political disputes between countries, and some of those disputes led to trade sanctions and political coercion outside the rules, but disputes were generally nested in the multilateral geopolitical order and were the exception, not the rule. Many could be resolved peacefully within the GATT framework. In that way international economic policy was largely siloed from national security policy. Within its ambit, the US hub-and-spokes security system added to political stability in an environment where multilateral trade rules could manage economic exchange to the benefit of countries that signed up to them.

That was then, while the rules could keep pace with developments in commerce and before the rise of China and other emerging countries meant the system could no longer be led mostly by one superpower, the United States. The system has steadily and fundamentally changed and is now under the biggest threat that it has faced since its creation in the ashes of world war.

The post war economic order has in part become a victim of its own success. The structure of global power has changed dramatically, due to the economic rise of Asia and especially China. China's entry to the WTO in 2001 was a watershed moment for the global trading system and helped China rise to become the world's largest trading nation and second largest economy within a decade. China's reform process and its transition from a centrally planned economy accelerated but in key areas like the relationship between the state and the market, with its state-owned enterprises, reform stalled. China's weight in the global economy means that the uneven playing field in China has significant spill-overs internationally. Not all countries see China's measures as an economic threat, and the world has benefited significantly from China's economic growth. But some advanced economies see forced technology transfer, industrial subsidies and intellectual property theft directly damaging their interests. Those are areas that existing WTO rules do not adequately address.

China has also deployed economic leverage and weaponized interdependence against other countries aimed at achieving political concessions, shaking the confidence of many countries in China's commitment to the spirit and rules of the multilateral trading system.

Multilateral trade rules have also failed to keep up with modern commerce. The WTO Doha Round of Development that was launched in 2001 failed to conclude and, until the 12th Ministerial Conference of the WTO in 2022 (MC12), there had been no multilateral agreements or significant progress in updating rules in the WTO. Rule-making in e-commerce and digital, intellectual property protections, labor and environmental standards and investment protection, as well as trade and investment liberalization has occurred in bilateral, regional and plurilateral rather than in multilateral agreements. The progress outside of the WTO framework is still built on the premise of WTO-plus agreements that rely on the underpinning of the WTO because of the uneven coverage of issues and countries across agreements, and reliance on dispute settlement in the WTO. The patchwork of rules from smaller agreements that try to cover these issues leave major gaps in trade governance and could cause economic fragmentation in the global economy.

Perhaps the biggest direct threat to the multilateral trading system has come from protectionism in the United States. Growing inequality in the United States that accelerated after the global financial crisis in 2008 and strategic competition with China have led to the United States actively undermining the WTO. The structural problems that led to the rise of President Trump in the United States — growth in inequality, the erosion of the social safety net and fracturing of the social compact, as well as a political psychology triggered by a relative decline in US global power — will take time to remedy.

The 'America First' protectionist policies of President Trump led to a series of unilateral trade measures aimed at other countries that resulted in a trade war with China and managed trade deals with Europe and Japan, that are anathema to free trade. The US veto of appellate judge appointments left the dispute settlement system in the WTO unable to enforce its rules for all members since 2019. By the end of 2022 there was no doubt that the Biden Administration in Washington had joined its predecessor the Trump administration on the mission to ignore the rules-based international trading system.

The Biden Administration is pursuing a foreign policy for the middle class and the domestic problems still drive US policy strategies and its foreign policy posture. US Trade Representative Katherine Tai left no doubt as to the Biden administration's rejection of global economic rules in her response to the December 2022 WTO ruling against the steel and aluminum tariffs that the United States deployed in the name of national security. Tai declared that the WTO "should not get into the business of second-guessing the national-security decisions that are made by sovereign governments".

Article XXI of the GATT, the security exceptions, had always allowed countries to impose restrictive measures for genuine national security reasons. However, the exemption was never intended to be a blanket one: there was an implicit agreement not to overstep the mark that stopped the rules of the GATT and then the WTO from being shredded in the name of national security. That norm holds no longer.

There is a growing overlap of interests between protectionist forces and the national security community in the United States that has justified large-scale industrial subsidies and extraterritorial sanctions in the name of security.

Most recently, in the US, the increasing focus on nation-ranking has manifested in the real possibility of losing primacy to a rising China. Adding to this, protectionism has seen its appeal powerfully ratcheted up by internal socioeconomic and political challenges. The Realist thinking that has emerged to cast payoffs in terms of relative gains, rather than absolute benefits, has therefore unhelpfully strengthened the tendency towards zero-sum perspective. These developments explain why Washington's approach to China has changed from engagement to a goal of containment.

China's trade policy responses too have been retaliatory. Coupled with China's increasingly strident nationalistic narrative, those actions have brought into sharp relief the zero-sum competition between China and the US. The behaviour of both have infected policy strategies across the rest of the world. Individual efforts to get ahead in a world perceived to be zero-sum runs the risk of a global descent into an epic-fail outcome in international economic policy last seen in the 1930s.

The Biden administration introduced the CHIPS and Science Act in late 2022, along with export controls that sought to limit Chinese participation in the complex international semiconductor chip trade and production networks. There is no longer any pretense of not forcing countries to choose between China and the United States — if US allies remain in the advanced semiconductor business with China, they will be hit by sanctions. The Biden administration also introduced the Inflation Reduction Act which creates an uneven playing field with the subsidization of electric vehicle manufacturing in the United States with large scale buy-in to industrial policy — exactly the issue it has been accusing the Chinese of — and retreat from free trade.

This is a significant U-turn in US economic policy and a major blow to the rules-based economic order, of which the United States had historically been the primary defender. It's a development of systemic importance because the US, however challenged its economic and social infrastructure, remains the largest economy in the world and the world's second largest trader. The United States is now less important than it used to be in the world economy and global trade, but it's still the world's superpower and its innovation and moral authority mean that countries still look to Washington to lead.

The United States and China, the world's two largest economies, are locked into strategic competition and rivalry that complicates international policy choices for the rest of the world, and particularly for their partners in Asia that are deeply integrated into both economies. Each Great Power sees any gain by the other side as a loss to themselves. Competition is, therefore, zero-sum in perception. But in substantive ways that rivalry is epic fail, in each side accepting loss to itself as long as the other side is disadvantaged. Global cooperation is impossible from the coming together of zero-sum mindsets and epic-fail outcomes.

What then, can the rest of the world — Third Nations — do in response to the realities of Great Power competition to further global cooperation?

## 5. Policy Implications: Third Nation Agency

### *Third Nation Agency as Facilitation for Inadvertent Cooperation*

Cooperation can bring about efficient win-win outcomes. But while cooperation is sufficient, it is not necessary. Mechanism design is a branch of economics that seeks to come up with incentive schemes so that individual incentive-advancing actions alone—without the need for explicit cooperation—will produce efficient outcomes. We call this **inadvertent cooperation**.

Our mechanism-design proposal begins with bringing other nations into the global discourse alongside the Great Powers. Third Nations—those whose people live in neither China nor the US—constitute 80% of humanity.  No Third Nation, by itself, would want to stand up against any of the Great Powers, but that is not what we intend. Such head-on confrontation is a simplistic view of what Third Nations need to do. Instead, what Third Nations need to do is provide nudges that dislodge the epic-fail outcome as an equilibrium.

Third Nations should set out the following:

1. Dispense with the idea that there is a top, number 1 nation in the world. Shoring up this exceptionalist, slightly out of date narrative only continues the unhelpful perception of a zero-sum world when blatantly many win-win positive outcomes are available, and in the case of avoiding climate change, necessary.  Instead, Third Nations can drive a narrative of there being multiple number 1 nations, each specific to a particular problem domain.
2. Identify problem domains where Third Nations can provide consequential agency. Sure, many challenges need the heft of a Great Power. Those are problems whose nature makes size disproportionally effective, i.e., problems whose solutions show the benefits of increasing returns to scale. Military confrontation is an example of such a problem, as are many global public goods and bads. Bottom-up rule-making by Third Nations can create solutions to Great Powers to avoid epic-fail equilibria.
3. Isolate and contain problem domains that are truly zero-sum. Nothing any Third Nation or even Great Power can do will improve outcomes there. Provided those zero-sum problems are then in a sufficiently small sandbox, all other issues have epic fail potential, and so need attention to nudge them towards win-win outcomes.
4. Abide by and reinforce multilateral rules and principles even when Great Powers deviate.  Acting collectively sets a "high-road" moral suasion example, but also helps retain most of the economics of scale benefits of multilateralism.

By design, 1.–4. are spaces where gentle nudges from Third Nations can critically exercise agency to move outcomes away from epic-fail equilibria. These measures can help produce conditions and frameworks to bring about inadvertent cooperation.[8]

Third Nations should navigate away from situations where Great Powers force them to choose sides. Economics teaches that choice expands the space of options; it does not restrict. Thus, saying Third Nations are free to choose but then reducing that decision to one of merely picking sides — this sphere of influence or that — sacrifices logic on the altar of national security. Conceiving choice as no more than alignment privileges the Great Powers at the expense of Third Nations. This is not to suggest naivete or military unpreparedness without Great Power protection. Far from it. But such a casting of Third Nations' choosing is nothing more than a surrender of agency. It subordinates the broad interests of Third Nations to the narrower ones of the Great Powers engaged in a zero-sum game of geopolitical rivalry.

---

[8] An alternative view on 1.-4. is that they might even incentivise unilateral action that delivers positive-sum outcomes.  Vines (2023) refers to this as **concerted unilateralism**.  This, in our analysis, is a striking and important special case of inadvertent cooperation.

There are instructive recent examples of Third Nation agency. One is the appeal to the multilateral system and WTO in the G7 Hiroshima Leaders' Communiqué as central to achieving economic security and resilience, likely driven by host Japan with support of Canada and European countries given the US stance towards the WTO. Europe's reframing of the US approach towards China from decoupling to de-risking is another. Those are examples from countries aligned with the United States that are acting to preserve their multilateral space and avoid narrowing of options. More will be needed from the non-aligned groupings such as ASEAN that are explicit about avoiding picking sides. Initiatives from ASEAN in trade in the form of the Regional Comprehensive Economic Partnership, and strategically in the form of its Outlook on the Indo-Pacific are expressions of multilateral and inclusive interests that need to be strengthened.

Third Nations can create the frameworks that shape engagement. These frameworks require two conditions: first, ideas that change the incentives for international engagement and second, groupings that mobilise collective action to implement those ideas.

In 1961 the so-called Non-Aligned Movement (in our interpretation, the then-counterpart of Third Nations) was formally established to counter the rising bipolarity of the Cold War. The movement drew on principles articulated at the Bandung 1955 Conference, including respect for national sovereignty and territorial integrity; recognition of equality across all races and all nations; settlement of international disputes through peaceful negotiations; respect for international obligations and justice; prohibiting nations exerting undue pressure on others to serve the interests of particular nations. A contemporary version, perhaps **Bandung 2.0**, might be timely as a channel by which Third Nations can explain the dangers of epic-fail outcomes due to Great Power rivalry.

A broader conception of security that incorporates economic, environmental, energy and human security objectives alongside military security can reframe outcomes away from vicious-cycle towards virtuous-cycle, away from epic-fail to mutual positive gain. Such a comprehensive security framing is not new in East Asia: it had been advocated strongly by Japan in the late 1970s and subsequently integrated into ASEAN thinking. On the other hand, political and economic fragmentation are the more likely outcomes from conventional security responses that unwind economic interdependence, prioritise military deterrence, and weaponise economic interdependence.

# 6. Conclusions

Around the world, multiple global crises have emerged coincident with the rise in every country of a national security imperative dominating the economic imperative. A defining feature of national security concerns is their focus on tail risks. Given the consequentiality of such concerns, policy formulation should give due weight to them. At the same time, however, given the risk profile underlying such concerns, neither the world nor the nation is necessarily made safer by allowing national security concerns to dominate policy thinking: Ignoring this is what makes for epic failure outcomes.

To improve performance, rules are needed in economic exchange, international markets, and economic governance that appropriately factor national security concerns into their operations. But open markets and transparent rules and norms can do much to help preserve interactions between nations and thus avoid epic-fail outcomes that make all poorer and, indeed, less secure.

In this paper we have described the economics of the international economic order, and drawn on the theoretical framing to understand current challenges in the global economic order. The view we develop shows that economic relations, while a critical component of world order, is too often under-weighted when held up against considerations of national security. Such under-weighting is not a statement of the authors' personal judgement. It refers to how the international system carries vicious-cycle dynamics that, with appropriate nudging, can be turned into virtuous-cycle outcomes. Unfortunately, even as perceptions around the world have become more zero-sum, Great Power rivalry has worsened the equilibrium into epic fail. In the paper we have also provided an empirical narrative in support of our theoretical framing.

To summarise, the concrete implications that emerge from our analysis are four-fold: (1) in the current situation, considerations of national security have inappropriately edged out those of economics and economic prosperity; (2) a zero-sum perspective has set in, and itself provides an obstacle to improve cross-nation relations; this opens the possibility of even worse, epic-fail equilibria; (3) a recalibration of views on security and economic factors is needed, along the lines of comprehensive security; (4) Third Nations—states that are not Great Powers—can be critical in this recalibration. This does not mean Third Nations should be vainly seeking co-equality with Great Powers. Instead, Third Nations should be looking for ways to facilitate inadvertent cooperation. Elevating comprehensive security would already be an important step forwards. An updated Bandung Declaration—a Bandung 2.0, as we call it in the text—would be another. Strengthening open, transparent and multilateral features of existing arrangements and bringing them together with new initiatives with appeal to positive-sum cooperation can guide collective action that helps avoid a global epic-fail.